%% file: preserverspanners.tex
\documentclass{article}

\usepackage[utf8]{inputenc}
\usepackage[left=3cm, right=3cm, top=2cm, bottom=2cm]{geometry}

\usepackage[linesnumbered,boxed]{algorithm2e}
\usepackage[noend]{algpseudocode}

\usepackage{amsmath}
\usepackage{amssymb}
\usepackage{amsthm}
\usepackage{authblk}
\usepackage{comment}
\usepackage{mdframed}

\usepackage{tikz}
\usetikzlibrary{patterns, decorations.pathreplacing}
\usepackage{subcaption}

\newtheorem{definition}{Definition}
\newtheorem{theorem}{Theorem}

\newtheorem{lemma}[theorem]{Lemma}

\newtheorem{conjecture}[theorem]{Conjecture}

\newcommand{\eps}{\varepsilon}

\newcommand{\eee}{\mathcal{E}}

\newcommand{\Oish}{\widetilde{O}}

\newcommand{\Ohat}{\widehat{O}}
\newcommand{\Thetahat}{\widehat{\Theta}}
\newcommand{\Omegahat}{\widehat{\Omega}}

\DeclareMathOperator{\dist}{dist}

\DeclareMathOperator{\poly}{poly}

\title{Better Distance Preservers and Additive Spanners\footnote{A preliminary version of this paper appeared in the conference SODA 2016.  Work performed while the authors were at Stanford University.}}
\author[1]{Greg Bodwin\thanks{bodwin@umich.edu} }
\author[2]{Virginia Vassilevska Williams\thanks{virgi@mit.edu}}
\affil[1]{University of Michigan}
\affil[2]{MIT EECS}
\date{}
\begin{document}

\maketitle

\thispagestyle{empty}
\begin{abstract}
\input{abstract}
\end{abstract}

\clearpage
\pagenumbering{arabic}

\input{intro}

\input{dps}

\input{addspan}

\section*{Acknowledgments}

We are grateful to several anonymous reviewers for useful comments and corrections that have improved this paper.

	\bibliographystyle{alpha}
	\bibliography{references}

\end{document}

%% file: abstract.tex
We study two popular ways to sketch the shortest path distances of an input graph.
The first is \emph{distance preservers}, which are sparse subgraphs that agree with the distances of the original graph on a given set of demand pairs.
Prior work on distance preservers has exploited only a simple structural property of shortest paths, called \emph{consistency}, stating that one can break shortest path ties such that no two paths intersect, split apart, and then intersect again later.
We prove that consistency alone is not enough to understand distance preservers, by showing both a lower bound on the power of consistency and a new general upper bound that polynomially surpasses it.
Specifically, our new upper bound is that any $p$ demand pairs in an $n$-node undirected unweighted graph have a distance preserver on $O(n^{2/3}p^{2/3} + np^{1/3})$ edges.
We leave a conjecture that the right bound is $O(n^{2/3}p^{2/3} + n)$ or better.

The second part of this paper leverages these distance preservers in a new construction of \emph{additive spanners}, which are subgraphs that preserve all pairwise distances up to an additive error function.
We give improved error bounds for spanners with relatively few edges; for example, we prove that all graphs have spanners on $O(n)$ edges with $+O(n^{3/7 + \eps})$ error.
Our construction can be viewed as an extension of the popular path-buying framework to clusters of larger radii.

%% file: intro.tex
\section{Introduction}

How much can graphs be compressed while keeping their distance information roughly intact?
This question falls within the scope of both metric embeddings and graph theory and is fundamental to our understanding of the metric properties of graphs.
When the compressed version of the graph must be a subgraph, it is called a {\em spanner}.
Spanners have a multitude of applications, essentially everywhere shortest path information needs to be compressed while still allowing for graph algorithms to be run, and they have been intensively studied theoretically \cite{PU89jacm, PU89sicomp, ADDJS93, ACIM99, DHZ00, EP04, TZ06, BKMP10, Chechik13soda, AB17jacm, FS16, ABP17}.
The quality of a spanner is measured by the tradeoff between its sparsity and its accuracy in preserving distances.
There are several different versions of spanners; we will discuss two of them below.

\subsection{Distance Preservers}

One possible version of the spanner problem is \emph{distance preservers}, in which we must preserve \emph{some} of the pairwise distances \emph{exactly}.
All graphs in this paper are undirected and unweighted.
\begin{definition} [Distance Preservers \cite{CE06}] \label{def:dps}
Given a graph $G = (V, E)$ and a set $P \subseteq V \times V$ of demand pairs, we say that a subgraph $H \subseteq G$ is a \emph{distance preserver} of $G, P$ if $\dist_H(u, v) = \dist_G(u, v)$ for all $(u, v) \in P$.
\end{definition}

Distance preservers are fundamental combinatorial objects: they generalize BFS trees (which are distance preservers for the special case $P = \{s\} \times V$), they have applications to many other types of graph sketches \cite{BCE05, Pettie09, BV15, EP16, AB16soda, AB17jacm, ABP17, CGMW18, HP18}, and they have been recently popular objects of study in their own right \cite{BCE05, CE06, Bodwin21sicomp, Bodwin19, CGMW18, CGK13, KV15, Kavitha17, GR17}.
Since a distance preserver must contain the edges of a shortest path for each demand pair, without loss of generality constructing a distance preserver is equivalent to choosing a \emph{tiebreaking scheme}, i.e.\ a way to select one of the possibly many shortest paths for each demand pair, and then overlaying the chosen shortest paths to obtain the preserver.
A basic property a tiebreaking scheme can have is \emph{consistency}, meaning that the intersection of any two chosen shortest paths $\pi_1, \pi_2$ is a single (possibly empty) continuous subpath of each.
One major reason consistency is fundamental is because it is the property that gives rise to BFS trees: that is, for demand pairs $\{s\} \times V$, one gets a distance-preserving tree if and only if one overlays shortest paths according to a consistent tiebreaking scheme.
Another reason consistency is fundamental is that it gives the current state-of-the-art upper bounds for distance preservers in general: given $p$ demand pairs in an $n$-node graph, a consistent tiebreaking scheme will yield a distance preserver on $O(\min\{n+n^{1/2}p, np^{1/2}\})$ edges \cite{CE06}.
This result is even partially tight in the following sense: it implies that $p = O(\sqrt{n})$ demand pairs have a distance preserver on $O(n)$ edges, and this corollary cannot possibly be improved by increasing the range to any $p = \omega(n^{1/2})$ while still paying only $O(n)$ edges \cite{CE06}.

Given all this, it is natural to ask if we actually need to investigate any deeper properties of shortest paths, or if consistency alone will suffice to understand distance preservers.
In this work, we answer this question with the following two new results:

\begin{theorem} \label{thm:badcons}
For any integer $p = O(n^2)$, there is an infinite family of $n$-node graphs, sets of $p$ demand pairs, and consistent tiebreaking schemes for which the resulting distance preserver has $\Omega(\min\{n + n^{1/2}p, np^{1/2}\})$ edges.
\end{theorem}

\begin{theorem} \label{thm:gooddps} Every $n$-node graph and set of $p$ demand pairs has a distance preserver on $O(n^{2/3}p^{2/3} + np^{1/3})$ edges.
\end{theorem}

In particular, Theorem \ref{thm:badcons} implies that the analysis in \cite{CE06} \emph{of consistent tiebreaking} was exactly tight, but Theorem \ref{thm:gooddps} shows that \emph{different tiebreaking} improves on it polynomially -- so consistency alone is not enough.
The $+np^{1/3}$ term in Theorem \ref{thm:gooddps} arises for purely technical reasons; it is unlikely to be really needed, but we have been unable to remove it.

\begin{conjecture} \label{cjt:greatdps}
Every $n$-node graph and set of $p$ demand pairs has a distance preserver on $O(n^{2/3}p^{2/3} + n)$ edges.
\end{conjecture}

\begin{figure}[ht]
\centering
\begin{tikzpicture}[xscale=0.5, yscale=0.5]
  	\draw (0,0) -- coordinate (x axis mid) (12,0);
    \draw (0,0) -- coordinate (y axis mid) (0,12);
    
    \foreach \x in {0,...,12}
    	\draw (\x,1pt) -- (\x,-5pt);
    \foreach \y in {0,...,12}
    	\draw (1pt,\y) -- (-5pt,\y);

	\node[rotate=90, above=0.2cm] at (0, 12) {$n^2$};
	\node[rotate=90, above=0.2cm] at (0, 9) {$n^{7/4}$};
	\node[rotate=90, above=0.2cm] at (0, 6) {$n^{3/2}$};
	\node[rotate=90, above=0.2cm] at (0, 3) {$n^{5/4}$};
	\node[rotate=90, above=0.2cm] at (0, 0) {$n$};
	
	\node[below=0.2cm] at (12, 0) {$n^2$};
	\node[below=0.2cm] at (9, 0) {$n^{3/2}$};
	\node[below=0.2cm] at (6, 0) {$n$};
	\node[below=0.2cm] at (3, 0) {$n^{1/2}$};
	\node[below=0.2cm] at (0, 0) {$1$};

	\node[below=0.8cm] at (x axis mid) {number of demand pairs $p$};
	\node[rotate=90, above=0.8cm] at (y axis mid) {edges in preserver $|E|$};
	
	\draw [fill=blue, ultra thick, blue] (12, 12) circle [radius=0.15];
	\draw [blue, thick] (6, 6) -- (12, 12);
	
	\draw [fill=blue, ultra thick, blue] (3, 0) circle [radius=0.15];
	\draw [blue, thick] (3, 0) -- (6, 6);
	
	\draw [fill=black!30!green, ultra thick, black!30!green] (6, 4) circle [radius=0.15];
	\draw [black!30!green, thick] (6, 4) -- (12, 12);
	\draw [black!30!green, thick] (6, 4) -- (0, 0);
	
	\draw [black!30!green, dotted, ultra thick] (6, 4) -- (3, 0);
	
	\draw [red, thick, dashed] (3, 0) to[bend right=20] (12, 12);
		
\end{tikzpicture}
\caption{\label{fig:dpub} The state-of-the-art asymptotic tradeoffs for distance preservers.
The upper bounds of \cite{CE06} are in solid blue, of Theorem \ref{thm:gooddps} in solid green, and of Conjecture \ref{cjt:greatdps} in dotted green.
The dashed red curve is, roughly, a lower bound from \cite{CE06}.
Some subpolynomial improvements in \cite{Bodwin21sicomp} are not pictured here.}
\end{figure}

\subsection{Additive Spanners}

Another type of spanner is when \emph{all} pairwise distances must be preserved \emph{up to an error term}:

\begin{definition} [Additive Spanners] \label{def:addspan}
Given a graph $G = (V, E)$, a subgraph $H \subseteq G$ is a {\em $+\beta$ (additive) spanner} of a graph $G$ if $\dist_H(u, v) \le \dist_G(u, v) + \beta$ for all $u, v \in V$.
\end{definition}

There are many interesting constructions with $\beta$ as an absolute constant: all $n$-node graphs have $+2$ spanners on $O(n^{3/2})$ edges \cite{ACIM99}, $+4$ spanners on $\Oish(n^{7/5})$ edges \cite{Chechik13soda}, and $+6$ spanners on $O(n^{4/3})$ edges \cite{BKMP10} (see also \cite{Knudsen14, Knudsen17, Woodruff10}).
But unfortunately, this setting then hits a dead-end: there is \emph{not} a general construction of $+\beta$ spanners on $O(n^{4/3 - \eps})$ edges, for any choice of absolute constants $\beta, \eps > 0$ \cite{AB17jacm} (see also \cite{Woodruff06, AB16soda, HP18}).
In particular, to obtain additive spanners in the regime $O(n^{4/3 - \eps})$, one essentially has to settle for $\beta = \poly(n)$.
Many applications call for spanners of linear size ($O(n)$ edges, or maybe a little more), so the additive error $\beta_0$ attainable by a spanner on $O(n)$ edges has attracted particular research interest.
Currently, it is known that
$$\Omega\left(n^{1/11}\right) \le \beta_0 \le \Oish\left(n^{1/2}\right)$$
where the upper bound is by prior work of the authors \cite{BV15} and the lower bound is by Lu \cite{Lu19} (see also \cite{BCE05, Pettie09, Chechik13soda}).
We improve polynomially on the upper bound:
\begin{theorem} \label{thm:addspan}
Assuming Conjecture \ref{cjt:greatdps}, for any $\eps > 0$ and $1 \le \eee \le n^{1/3}$, every $n$-node graph has a $+O(n^{3/7 + \eps} \eee^{-9/7})$ additive spanner on $O_{\eps}(n \eee)$ edges.
Moreover, this result holds unconditionally when $\eee = O(1)$ or $\eee = \Omega(n^{2/13})$.
\end{theorem}
In particular, the new (unconditional) upper bound is
$\beta_0 \le n^{3/7 + \eps}.$

\paragraph{Technical Overview of Spanner Construction.} Here we give a brief technical overview of our spanner construction, aimed at a reader with some familiarity with the area.
Theorem \ref{thm:addspan} is based in part on the popular path-buying framework used in many previous spanner constructions, introduced in \cite{BKMP10}.
For background, these constructions begin with a \emph{clustering phase}, in which (1) we add all edges incident on ``low-degree'' nodes to the spanner, and (2) the ``high-degree'' nodes are partitioned into clusters of radius $1$ each.
Then they have a \emph{path-buying phase}, where some shortest paths are added to the graph with the goal of connecting the existing clusters.

To obtain even sparser spanners at the cost of more error, it is a natural idea to try to apply the same framework, but using clusters of radius $r \gg 1$.
Our key idea is that distance preservers play a key role in the proper extension of this method.
Instead of differentiating between \emph{high-degree} and \emph{low-degree} nodes, we propose an analogous dichotomy between high-radius clusters that are \emph{large} -- meaning that they contain many nodes -- and clusters that are \emph{bottlenecked}, meaning they can be separated from the rest of the graph by removing a relatively small set of nodes.
The idea nearly reduces to the high-degree/low-degree dichotomy in the extreme case of our construction where $r=1$, but it gains efficiency for larger $r$.
It turns out that bottlenecked clusters, analogous to low-degree nodes, can be handled cleanly with an application of distance preservers: one only has to add a distance preserver between pairs of nodes in the separating node set, and then for any shortest path, a subpath passing in and out of a bottlenecked cluster will already have all of its edges contained in a spanner. 

Theorem \ref{thm:addspan} is obtained by plugging the distance preservers of Conjecture \ref{cjt:greatdps} into this large/bottlenecked framework, and then proceeding through a reasonable generalization of the path-buying method of \cite{BKMP10} with a few extra optimizations.
Conjecture \ref{cjt:greatdps} is confirmed in the settings $p = O(n^{1/2})$ and $p = \Omega(n)$, and this corresponds to the ranges $\eee = O(1)$ and $\eee = \Omega(n^{2/13})$ in which Theorem \ref{thm:addspan} holds unconditionally.
One can also obtain a weaker yet unconditional version of Theorem \ref{thm:addspan} in the regime $\omega(1) \le \eee \le o(n^{2/13})$  by plugging the unconditional distance preserver bounds known in the intermediate regime.
These bounds are given in Section \ref{sec:uncond}.

%
%
%
\begin{figure}[ht]
\centering
\begin{tikzpicture}[scale=0.5]
  	\draw (0,0) -- coordinate (x axis mid) (12, 0);
    \draw (0,0) -- coordinate (y axis mid) (0,12);
    
    \foreach \x in {0,...,12}
    	\draw (\x,1pt) -- (\x,-5pt);
    \foreach \y in {0,...,12}
    	\draw (1pt,\y) -- (-5pt,\y);

	\node[below=0.2cm] at (12, 0) {$n^{3/2}$};
	\node[below=0.2cm] at (9, 0) {$n^{11/8}$};
	\node[below=0.2cm] at (6, 0) {$n^{5/4}$};
	\node[below=0.2cm] at (3, 0) {$n^{9/8}$};
	\node[below=0.2cm] at (0, 0) {$n$};
	
	\node[rotate=90, above=0.2cm] at (0, 12) {$n^{1/2}$};
	\node[rotate=90, above=0.2cm] at (0, 9) {$n^{3/8}$};
	\node[rotate=90, above=0.2cm] at (0, 6) {$n^{1/4}$};
	\node[rotate=90, above=0.2cm] at (0, 3) {$n^{1/8}$};
	\node[rotate=90, above=0.2cm] at (0, 0) {$1$};

	\node[below=0.8cm] at (x axis mid) {edges in spanner $|E|$};
	\node[rotate=90, above=0.8cm] at (y axis mid) {spanner error $+\beta$};
	
	\draw [ultra thick, fill=blue, blue] ({3/17*2*12}, {12 - 3/2*3/17*2*12}) circle [radius=0.2cm];
	\draw[ultra thick, blue] ({3/17*2*12}, {12 - 3/2*3/17*2*12}) -- (8, 0);
	\draw[ultra thick, blue] (0, 12) -- (3.3, 10.45);
	\draw[ultra thick, blue] ({3/17*2*12}, 8.9) -- (3.3, 10.45);
	\draw [ultra thick, blue, fill=white] ({3/17*2*12}, 8.9) circle [radius=0.2cm];
	
	\draw[ultra thick, black!30!green, dotted] (0, {3/7*2*12}) -- ({1/3*2*12}, 0);
	\draw[ultra thick, black!30!green] ({8*2/13*3}, {3/13*2*12}) -- (8, 0);
	\draw [fill=black!30!green, ultra thick, black!30!green] (0, {3/7*2*12}) circle [radius=0.2cm];
	
	\draw [dashed, red, ultra thick] plot [smooth] coordinates {(0, {2/11*12}) (1, 0.5)  (8, 0)};
\end{tikzpicture}
\caption{\label{fig:addub} The state-of-the-art asymptotic tradeoffs for additive spanners.
The upper part of the solid blue curve is an upper bound of \cite{BV15}, the lower solid blue line is the upper bound of \cite{Chechik13soda}, and the upper bound of Theorem \ref{thm:addspan} is in solid green, dotted in the regime where it is conditional on Conjecture \ref{cjt:greatdps}.
The dashed red line is, roughly, a lower bound from \cite{Lu19}.}
\end{figure}

%% file: dps.tex
\section{Distance Preservers}

We will be concerned only with existential distance preserver bounds in this paper and not with construction runtime.
We note that any existential bound is automatically algorithmic: that is, if $e^*(n, p)$ is the maximum number of edges needed for a distance preserver of $p$ demand pairs in an $n$-node graph, then there is a folklore algorithm that runs in polynomial time and always outputs a distance preserver on $\le e^*(n, p)$ edges.
This algorithm is simply to consider the edges of $G$ in any order and delete the current edge if doing so does not change the distance between any given demand pair.
By construction, the final graph $H$ will be the unique distance preserver of itself, so it cannot have more than $e^*(n, p)$ edges.

As discussed in the introduction, it is nice theoretically to reduce the construction of distance preservers to \emph{shortest path tiebreaking}:
\begin{definition} [Tiebreaking Schemes]
A \emph{tiebreaking scheme} on a graph $G$ is a map $\pi$ from node pairs $(u, v)$ to a shortest $u \leadsto v$ path $\pi(u, v)$ in $G$.
\end{definition}
\noindent A subgraph is a distance preserver iff it contains the edges of a shortest path for each demand pair, so w.l.o.g.\ we may assume that any distance preserver is constructed by selecting a tiebreaking scheme $\pi$ and then overlaying the chosen shortest paths for each demand pair.
In other words, given a graph $G$ and demand pairs $P$, the distance preserver $H$ associated to a given tiebreaking scheme $\pi$ is the one where an edge $e$ is included in $H$ iff there is $(u, v) \in P$ with $e \in \pi(u, v)$.

\subsection{Limitations of Consistent Tiebreaking}

\begin{definition} [Consistency]
A tiebreaking scheme $\pi$ is \emph{consistent} if, for any nodes $w, x, y, z$, we have that $\pi(w, x) \cap \pi(y, z)$ is a (possibly empty) continuous subpath of both $\pi(w, x)$ and $\pi(y, z)$.
\end{definition}
\noindent It is folklore that every graph has a consistent tiebreaking scheme.
For example, one is obtained by \emph{random reweighting}, in which we add a tiny random variable to each edge weight to break ties with probability $1$.
In the reweighted graph $G'$ all shortest paths are unique, and thus there is only one tiebreaking scheme $\pi$ over $G'$.
For any two paths $\pi(w, x), \pi(y, z)$ their intersection (if nonempty) is the unique shortest path between its endpoints, which is thus a continuous subpath of both $\pi(w, x)$ and $\pi(y, z)$.
Hence $\pi$ is consistent in $G'$, and so it is also consistent in $G$ when the original unit edge weights are restored.
Coppersmith and Elkin proved:

\begin{theorem} [\cite{CE06}]
For any $p$ demand pairs and consistent tiebreaking scheme $\pi$ in an $n$-node graph, the associated distance preserver has $O(\min\{n^{1/2} p + n, np^{1/2}\})$ edges.
\end{theorem}

Our first new result is that this is tight for consistency.
In particular:
\begin{theorem} [See Figure \ref{fig:badcons}] \label{thm:badconsbody}
For any $p = O(n^2)$, there is an infinite family of $n$-node graphs $G$, consistent tiebreaking schemes $\pi$, and $p$ demand pairs so that the associated distance preserver has $\Omega(\min\{n^{1/2}p + n, np^{1/2}\})$ edges.
\end{theorem}
\begin{proof}
Let $q$ be a prime and let $G$ be the graph whose nodes are the elements of the field $F_q^2$ and whose edge set contains exactly the node pairs whose first coordinate differs by $1$, not including ``wraparound'' edges of the form $((0, x), (q-1, y))$.
The demand pairs $P$ are exactly the node pairs $((0, x), (q-1, y))$ for any $x, y \in F_q$.
By a \emph{line} in $F_q^2$ we will mean a set of points that can be described by
$$\left\{(x, mx+b) \ \mid \ x \in F_q\right\}$$
for some fixed $m, b \in F_q$ (note this does not include ``vertical'' lines with fixed first coordinate).
Each demand pair $(u, v)$ uniquely determines a line in $F_q^2$ passing through both points.
Define $\pi(u, v)$ as the path containing exactly the nodes on this line.
Note that $\dist(u, v) = q-1$ since the first coordinate of two nodes only differs by $1$ over an edge; thus, each $\pi(u, v)$ is indeed a shortest path since it has exactly $q-1$ edges, so $\pi$ is a tiebreaking scheme.
Additionally, $\pi$ is consistent, since the paths $\pi(u, v)$ correspond to lines in $F_q^2$ and any two lines intersect on only $0$ or $1$ nodes.

The distance preserver associated to $\pi$ is $G$ itself, since every edge of $G$ similarly determines a line in $F_q^2$ which is thus equal to $\pi(u, v)$ for exactly one demand pair $(u, v)$. The graph $G$ has $q^2 =: n$ nodes and $(q-1)q^2 = \Theta(n^{3/2})$ edges, and we have $|P| = q^2 = n$ demand pairs, which gives our claimed lower bound in the special case when $p=n$.
To obtain our lower bound for other values of $p$:
\begin{itemize}
\item In the regime $n \le p \le n^2$, we modify our construction by taking only its first few layers.
That is: letting $k \le q-1$ be a parameter of the construction, we only let the first coordinate range from $0$ to $k$ but everything else remains the same (essentially, the nodes of $G$ are now $[k] \times F_q$).
This gives a lower bound on $n := kq$ nodes, $(k-1)q^2 = \Theta(n^2/k)$ edges, and $q^2 = n^2/k^2$ demand pairs.
The lower bound thus follows by choosing $k = \Theta(n/\sqrt{p})$.

\item In the regime $\sqrt{n} \le p \le n$, we modify our construction by arbitrarily choosing $p$ demand pairs to keep.
We then discard the rest, as well as all the edges in $G$ that appear in $\pi(s, t)$ for any discarded demand pair $(s, t)$.
The resulting construction has $n$ nodes, $p$ pairs, and since each surviving demand pair has $q - 1 = \Theta(\sqrt{n})$ distinct edges in its path, the lower bound has $\Theta(p\sqrt{n})$ edges.
\end{itemize}
Finally, the lower bound of $\Omega(n)$ in the regime $p \le \sqrt{n}$ is trivial (let $G$ be a big long path, and take the endpoints of this path as one of the demand pairs).
\end{proof}
\begin{figure}[ht]
\centering
\begin{tikzpicture}[scale=0.7]
\foreach \x in {0,...,6}
   \foreach \y in {0,...,6}
    	\draw [fill=black] (2*\x, \y) circle [radius=0.15];
    		
\node at (-0.4, -0.5) {$(0, 0)$};
\node at (-0.4, 6.5) {$(0, 6)$};
\node at (12.4, 6.5) {$(6, 6)$};
\node at (12.4, -0.5){$(6, 0)$};
	
\draw [red, ultra thick] (0, 6) -- (12, 6);

\draw [orange, ultra thick] (0, 6) -- (12, 0);

\draw [black!30!green, ultra thick] (0, 6) -- (6, 0);
\draw [black!30!green, ultra thick] (6, 0) -- (8, 5);
\draw [black!30!green, ultra thick] (8, 5) -- (12, 1);

\draw [blue, ultra thick] (0, 6) -- (4, 0);
\draw [blue, ultra thick] (4, 0) -- (6, 4);
\draw [blue, ultra thick] (6, 4) -- (8, 1);
\draw [blue, ultra thick] (8, 1) -- (10, 5);
\draw [blue, ultra thick] (10, 5) -- (12, 2);
\end{tikzpicture}
\caption{\label{fig:badcons} The construction described in Theorem \ref{thm:badconsbody}, with $q = 7$.  For clarity, only the first $4$ out of the $49$ shortest paths $\pi(u, v)$ are pictured here.}
\end{figure}

\subsection{Beyond Consistent Tiebreaking}

We will next prove our new upper bound on distance preservers.
We first develop two technical ingredients.
In the sequel it will be convenient to interpret demand pairs $(s, t) \in P$ to be ordered (e.g.\ $(s, t)$ rather than $(t, s)$), and we will write $(u, v) \in \pi(s, t)$ to mean that the nodes $u, v$ appear adjacently in that order in $\pi(s, t)$; that is, this notation implies that $u$ is closer to $s$ and $v$ is closer to $t$.
\begin{definition} [Branching Events \cite{CE06}]
Given a graph $G$, a set of demand pairs $P$, and a tiebreaking scheme $\pi$, a \emph{branching event} is a pair of distinct edges $(u, v), (w, v)$ that share an endpoint node $v$, such that there are two distinct demand pairs $p, p' \in P$ with $(u, v) \in \pi(p) \setminus \pi(p')$ and $(w, v) \in \pi(p') \setminus \pi(p)$.
\end{definition}

Due to our orientation of edges, this technically means (for example) that a directed tree rooted at a node $s$ could have $0$ branching events if the tree paths are all directed away from $s$, or $\Omega(n^2)$ branching events if the paths are directed towards $s$.
So branching events depend on the ordering of demand pairs ($(s, t)$ vs.\ $(t, s)$).
One could just as easily define branching events to be two oriented edges leaving the same node ($(v, u), (v, w)$), as opposed to two oriented edges entering the same node as we've done here; this choice is somewhat arbitrary.

\begin{lemma} [\cite{CE06}] \label{lem:brev}
If a tiebreaking scheme $\pi$ for $G, P$ has $b$ branching events, then the associated distance preserver has $O((nb)^{1/2} + n)$ edges.
\end{lemma}
\begin{proof}
Let $ H = (V, E)$ be the associated distance preserver.
Pick an arbitrary total ordering of the set $P$ and label each edge $e \in E$ with the first demand pair $(s, t) \in P$ in the ordering so that $e \in \pi(s, t)$.
Since we assign $e$ the orientation from $s$ towards $t$, we can say that each node $v \in V$ is the endpoint for $\ge {\deg_{in}(v) \choose 2}$ different branching events.
We then compute:
\begin{align*}
|E| &= \sum \limits_{v \in V} \deg_{in}(v)\\
&= O(n) + \sum \limits_{\left\{v \in V \ \mid \ \deg_{in}(v) \ge 2\right\}} \deg_{in}(v)\\
&= O(n) + O\left( \left( n \cdot \sum \limits_{\left\{v \in V \ \mid \ \deg_{in}(v) \ge 2\right\}} \deg_{in}(v)^2\right)^{1/2} \right) \tag*{Cauchy-Schwarz}\\
&= O(n) + O\left( \left( n \cdot \sum \limits_{v \in V} {\deg_{in}(v) \choose 2}\right)^{1/2} \right)\\
&= O(n) + O\left((nb)^{1/2}\right). \tag*{\qedhere}
\end{align*}
\end{proof}

\begin{lemma} [See Figure \ref{fig:reroute}] \label{lem:chokepres}
Let $G$ be an $n$-node graph and let $S$ be a subset of $|S|=\sigma$ nodes of diameter $d$ (meaning for all $s_1, s_2 \in S$ we have $\dist(s_1, s_2) \le d$).
Let $P \subseteq S \times V$ be a set of $|P| = p$ demand pairs.
Then there is a distance preserver of $G, P$ on
$$O\left((np\sigma d)^{1/2} + n\right)$$
edges.
\end{lemma}
\begin{proof}
Let $\pi$ be a tiebreaking scheme whose associated distance preserver $H$ has as few edges as possible.
Every edge $e$ can then be labeled with a demand pair $(s, t) \in P$ so that every $s \leadsto t$ shortest path in $H$ includes $e$ (if no such demand pair $(s, t)$ exists, then we can modify $\pi$ to make all paths avoid $e$, thus removing $e$ from $H$).
Each branching event $(e, e')$ is then labeled with a pair of demand pairs $p, p' \in P$, inheriting the labels of $e, e'$.

We will prove this lemma by showing that there are only $O(p\sigma d)$ total branching events in $H$; the lemma then follows by plugging $b = O(p\sigma d)$ into Lemma \ref{lem:brev}.
Suppose towards contradiction that we have $>p\sigma (2d+1)$ branching events in $H$; then by the pigeonhole principle there is a node $w \in S$, a demand pair $(x, y) \in P$, and (at least) $2d+2$ branching events $\{b_1, \dots, b_{2d+2}\}$ so that each branching event $b_i$ has a label of the form $(x, y), (w, z_i)$ (i.e.\ $x, y, w$ are the same in the labels of all these branching events, and only $z_i$ varies).
Let $v_i$ be the node shared by the two edges in the branching event $b_i$, and we will assume for convenience that the nodes $\{v_i\}$ are ordered by increasing $\dist(x, v_i)$.
By the triangle inequality and the diameter of $S$, for any $1 \le i \le 2d+2$ we have
$$d \ge \dist(w, x) \ge \dist(w, v_i) - \dist(x, v_i) \ge - \dist(w, x) \ge -d.$$
Hence, by the pigeonhole principle, there are $1 \le i < j \le 2d + 2$ with
$$\dist(w, v_i) - \dist(x, v_i) = \dist(w, v_j) - \dist(x, v_j)$$
and so
$$\dist(w, v_i) + \dist(x, v_j) - \dist(x, v_i) = \dist(w, v_j).$$
Note that $x, v_i, v_j$ all lie on $\pi(x, y)$ in that order, and so $\dist(x, v_j) - \dist(x, v_i) = \dist(v_i, v_j)$.
Therefore
$$\dist(w, v_i) + \dist(v_i, v_j) = \dist(w, v_j).$$
This equation implies that $v_i$ lies on a shortest $w \leadsto v_j$ path, and so we can form a shortest $w \leadsto v_j$ path by concatenating $\pi(w, v_i)$ with $\pi(x, y)[v_i \leadsto v_j]$.
This concatenated path enters $v_j$ along an edge in $\pi(x, y)$.
Thus, there exists a shortest $w \leadsto v_i$ path that uses an edge in $\pi(x, y)$ to enter $v_i$.
Hence it is \emph{not} the case that every shortest $w \leadsto v_i$ path includes an edge $e$ used in the branching event $b_i$.
This contradicts the labeling of our branching events.
\end{proof}

\begin{center}
\begin{figure}[ht]
\begin{minipage}{0.4\textwidth}
\begin{tikzpicture}[scale=0.7, xscale=0.7]
\draw [fill=black] (0, 3) circle [radius=0.15];
\node [above=0.2cm] at (0, 3) {$\mathbf{w}$};

\draw [fill=black] (0, 0) circle [radius=0.15];
\node [below=0.3cm] at (0, 0) {$\mathbf{x}$};

\draw [ultra thick, ->] (0, 0) -- (12, 0);
\draw [fill=black] (12, 0) circle [radius=0.15];
\node at (12, -0.5) {$\mathbf{y}$};

\draw [dashed] (0, 3) -- (0, 0);
\node at (-0.3, 1.5) {$d$};

\draw [fill=black] (2, 0) circle [radius=0.15];
\node [below=0.3cm] at (2, 0) {$\mathbf{v_1}$};
\draw [fill=black] (4, 0) circle [radius=0.15];
\node [below=0.3cm] at (4, 0) {$\mathbf{v_2}$};

\node [above = 0.2cm] at (6, 0) {$\mathbf{\dots}$};
\node [below = 0.2cm] at (6, 0) {$\mathbf{\dots}$};

\draw [fill=black] (8, 0) circle [radius=0.15];
\node [below=0.3cm] at (8, 0) {$\mathbf{v_{2d+1}}$};
\draw [fill=black] (10, 0) circle [radius=0.15];
\node [below=0.3cm] at (10, 0) {$\mathbf{v_{2d+2}}$};

\draw [ultra thick, red] (0, 3) to[bend left] (2, 0);
\draw [ultra thick, orange] (0, 3) to[bend left] (4, 0);
\draw [ultra thick, black!30!green] (0, 3) to[bend left] (8, 0);
\draw [ultra thick, blue] (0, 3) to[bend left] (10, 0);

\end{tikzpicture}
\subcaption{If $\pi$ routes $2d+2$ paths for demand pairs starting with $w$ to branch with another path $\pi(x, y)$ ...}
\end{minipage}%
\hspace{0.2\textwidth}%
\begin{minipage}{0.4\textwidth}
\begin{tikzpicture}[scale=0.7, xscale=0.7]
\draw [fill=black] (0, 3) circle [radius=0.15];
\node [above=0.2cm] at (0, 3) {$\mathbf{w}$};

\draw [fill=black] (0, 0) circle [radius=0.15];
\node [below=0.3cm] at (0, 0) {$\mathbf{x}$};

\draw [ultra thick, ->] (0, 0) -- (12, 0);
\draw [fill=black] (12, 0) circle [radius=0.15];
\node at (12, -0.5) {$\mathbf{y}$};

\draw [dashed] (0, 3) -- (0, 0);
\node at (-0.3, 1.5) {$d$};

\draw [fill=black] (2, 0) circle [radius=0.15];
\node [below=0.3cm] at (2, 0) {$\mathbf{v_1}$};
\draw [fill=black] (4, 0) circle [radius=0.15];
\node [below=0.3cm] at (4, 0) {$\mathbf{v_2}$};

\node [above = 0.2cm] at (6, 0) {$\mathbf{\dots}$};
\node [below = 0.2cm] at (6, 0) {$\mathbf{\dots}$};

\draw [fill=black] (8, 0) circle [radius=0.15];
\node [below=0.3cm] at (8, 0) {$\mathbf{v_{2d+1}}$};
\draw [fill=black] (10, 0) circle [radius=0.15];
\node [below=0.3cm] at (10, 0) {$\mathbf{v_{2d+2}}$};

\draw [ultra thick, red] (0, 3) to[bend left] (2, 0);
\draw [ultra thick, orange] (0, 3) to[bend left] (4, 0);
\draw [black!30!green, ultra thick] plot [smooth] coordinates {(0, 3) (2.3, 2.3) (4.2, 0.2) (8, 0)};
\draw [ultra thick, blue] (0, 3) to[bend left] (10, 0);

\end{tikzpicture}

\subcaption{... then there is an alternate shortest $w \leadsto v_j$ path for one of these nodes, which uses a different edge entering $v_j$.}
\end{minipage}
\caption{\label{fig:reroute} The pigeonhole technique used in Lemma \ref{lem:chokepres}.}
\end{figure}
\end{center}

In order to convert this lemma to an upper bound on distance preservers in general, we will use the following standard lemma in graph theory:
\begin{lemma} \label{lem:densenode}
Let $G = (V, E)$ be a nonempty graph with average degree $\eee$.
Then there exists a node $v$ that has $\Omega(\eee)$ neighbors of degree $\Omega(\eee)$ each.
\end{lemma}
\begin{proof}
While there exists a node of degree $\le \eee/4$, remove that node and all of its incident edges from $G$.
Letting $n$ be the initial number of nodes in $G$, the total number of edges removed in this way is $\le n\eee /4 \le |E|/2$, so when this process halts $G$ still has at least $|E|/2$ edges, so it is nonempty and it has minimum degree $\Omega(\eee)$.
So we can pick any node in the remaining graph.
\end{proof}

\begin{theorem}\label{thm:goodddps}
Every $n$-node graph and set of $|P|=p$ demand pairs has a distance preserver on $O(n^{2/3} p^{2/3} + np^{1/3})$ edges.
\end{theorem}
\begin{proof}
Let $\eee$ be an integer parameter that we will choose later.
Let $\pi'$ be any tiebreaking scheme and let $H' = \pi'(P)$ be the associated distance preserver.
Repeat the following process until the average degree in $H'$ is $\le \eee$.

Using Lemma \ref{lem:densenode}, we can find a node a node $v$ in $H'$ and a set $S$ containing $|S| = \Omega(\eee)$ neighbors of $v$ which all have degree $\Omega(\eee)$.
Hence there is a set $Q \subseteq P$ of $|Q| = \eee^2$ demand pairs so that, for every $(x, y) \in Q$, the path $\pi'(x, y)$ contains a node in $S$.
Letting $s_{x, y} \in S$ be a node so that $s_{x, y} \in \pi'(x, y)$, we may split each demand pair $(x, y)$ into two demand pairs $(s_{x, y}, x), (s_{x, y}, y)$.
Now, using Lemma \ref{lem:chokepres} (with parameters $d=2, \sigma \le \eee, p = 2\eee^2$) we can build a distance preserver of the demand pairs in $Q$ on
$$O\left( \sqrt{n |Q| |S|} + n \right) = O\left(n^{1/2} \eee^{3/2} + n\right)$$
edges.

We can now remove the (original) demand pairs in $Q$ from $P$, and repeat.
Since we remove $\Theta(\eee^2)$ demand pairs from $P$ in each round, we repeat only $O(p/\eee^2)$ times in total.
Thus, if we union together the distance preservers computed in each round and also the final leftover preserver $H'$, the total size is
$$\underbrace{O\left(\frac{p}{\eee^2}\right)}_{\text{number of rounds}} \cdot \underbrace{O\left(n^{1/2} \eee^{3/2} + n\right)}_{\text{size in each round}} + \underbrace{O(n \eee)}_{\text{size of } H'} = O\left(pn^{1/2} \eee^{-1/2} + np \eee^{-2} + n \eee\right)$$
edges.
Choosing $\eee = p^{2/3} n^{-1/3}$ gives
$$|E| = O\left(n^{2/3} p^{2/3} + n^{5/3} p^{-1/3}\right),$$
which is $O(n^{2/3}p^{2/3})$ when $p = \Omega(n)$.
Choosing $\eee = p^{1/3}$ gives
$$|E| = O\left( np^{1/3} + n^{1/2} p^{5/6}\right),$$
which is $O(np^{1/3})$ when $p = O(n)$.
\end{proof}

It seems that the $+np^{1/3}$ term arises in this argument for a purely technical reason.
The culprit is essentially the $+n$ in Lemma \ref{lem:brev}, which is amplified over the union bound of the separate preservers built for each set $Q$ in the above proof.
Of course the $+n$ in Lemma \ref{lem:brev} can't be removed in general, as shown by the counterexample where $G$ is a big long path, which needs $\Omega(n)$ edges for a distance preserver despite having $b=0$ branching events.
However, such graphs generally admit very good distance preservers when many demand pairs are considered.
We thus conjecture that with more care this problem can be avoided.
\begin{conjecture} \label{cjt:greatddps}
Every $n$-node graph and $p$ demand pairs have a distance preserver on $O(n^{2/3} p^{2/3} + n)$ edges.
\end{conjecture}

%% file: addspan.tex
\section{Additive Spanners}

We now give our additive spanner construction.
As discussed previously, our bounds are parametrized on an arbitrary $\eps > 0$.
We will use standard $O_{\eps}(\cdot)$ notation to hide multiplicative factors that depend only on the choice of $\eps$ (but not on $n$), and we will use notation $\Ohat(\cdot)$ to hide factors of the form $n^{O(\eps)}$.
(We also use $\Thetahat, \Omegahat$ notation defined analogously.)
We will prove:
\begin{theorem} \label{thm:adddspan}
Assuming Conjecture \ref{cjt:greatddps}, for any $\eps > 0$ and $1 \le \eee \le n^{1/3}$, every $n$-node graph $G = (V, E)$ has an additive spanner with
$$+ \Ohat\left(n^{3/7} \eee^{-9/7}\right)$$
error and $O_{\eps}(n \eee)$ edges.
\end{theorem}

At the end, we will also give versions of this theorem with slightly worse error bounds that hold unconditionally.
We let $\beta$ be a parameter of the construction, which roughly controls the spanner error (in the end we will set $\beta \approx n^{3/7} \eee^{-9/7}$).
We will use the following notations:
\begin{align*}
B(v, r) &:= \left\{u \in V \ \mid \ \dist(u, v) \le r\right\}, \text{ and}\\
B_{=}(v, r) &:= \left\{u \in V \ \mid \ \dist(u, v) = r\right\}.
\end{align*}

We view our approach as an extension of the classic path-buying method from \cite{BKMP10} to clusters of larger radius; our construction reduces to something very similar to the one in \cite{BKMP10} when the parameters are set such that $\beta = O(1)$.
Although familiarity with \cite{BKMP10} is not required to read the following proofs, it may be helpful, and we will occasionally pause to explain analogs to this argument in \cite{BKMP10} for the reader who is familiar with this prior work.

\subsection{Clustering}

The construction in \cite{BKMP10} begins with a \emph{clustering}, in which the nodes of the graph are either ``low-degree'' or ``high-degree;'' they accept all edges incident to low-degree nodes into the spanner, and the high-degree nodes are covered by a small number of clusters of diameter $2$ each.
In our setting where $\text{poly}(n)$ spanner error is allowed (\cite{BKMP10} only allows $+6$ error), it is natural to attempt an analogous method, using clusters of radius $\text{poly}(n)$ instead of $2$.
We set up an analogous clustering in the following two lemmas.
Our ``bottlenecked clusters'' are analogous to the low-degree nodes from \cite{BKMP10}, while the ``large clusters'' are analogous to the clusters that cover high-degree nodes.

\begin{lemma}  \label{lem:basecluster}
There exists a set of $k$ nodes $\{v_1, \dots, v_k\}$ and $k$ associated integers $\{r_1, \dots, r_k\}$, where each $r_i = \Thetahat(\beta)$, satisfying the following two properties:
\begin{itemize}
\item (Coverage) For each $v \in V$, there is $1 \le i \le k$ such that $v \in B(v_i, r_i)$.
\item (Non-Overlap) $\sum \limits_{i=1}^k \left|B(v_i, 2r_i)\right| = O_{\eps}(n)$.
\end{itemize}
(Note the different radii, $r_i$ vs.\ $2r_i$, in the coverage/non-overlap properties.)
\end{lemma}
\begin{proof}
First, for every node $v \in V$ we compute a \emph{potential radius} $r'_v$ as follows.
This will ultimately be half the radius of the cluster centered at $v$, if $v$ is chosen as a cluster center.
Let $c$ be a parameter and initialize $r'_v \gets \beta$.
If
$$c\left|B(v, r'_v)\right| \ge \left|B(v, 4r'_v)\right|,$$
then finalize $r'_v$ at its current value.
Otherwise, set $r'_v \gets 4r'_v$ and repeat.
Since $|B(v, r'_v)| \le n$ always, and this quantity grows by at least $\cdot c$ in each iteration, we repeat $\le \log_c n$ times.
Thus we have
$$r'_v \le \beta 4^{\log_{c} n} = \beta n^{\frac{1}{\log_4 c}}.$$
We choose $c = \Theta(4^{1/\eps})$, and so $r'_v \le \beta n^{\eps}$.
Our next step is to select the nodes that will be cluster centers.
Sort the nodes $v \in V$ descendingly by value of $r'_v$.
We repeat the following process until we are out of nodes.
Remove the first remaining node $v$ from the list and add it to the list of selected cluster centers.
Then, for each remaining node $u$ in the list with $B(u, r'_u) \cap B(v, r'_v) \ne \emptyset$, delete $u$ from the list.

The final radius associated with each selected cluster center $u$ will be $r_u := 2r'_u$.
To verify the coverage property: let $v \in V$.
If $v$ is chosen as a cluster center we immediately have $\dist(v, v) = 0 \le 2r'_v$.
Otherwise, we discarded $v$ during the construction, so there is a cluster center $u$ considered before $v$ with $B(u, r'_u) \cap B(v, r'_v) \ne \emptyset$.
Since $r'_u \ge r'_v$, this implies $\dist(v, u) \le 2r'_u = 2r_u$.
For the non-overlap property, we compute:
\begin{align*}
\sum \limits_{i=1}^k \left|B(v_i, 2r_i)\right| &= \sum \limits_{i=1}^k \left|B(v_i, 4r'_i)\right|\\
&\le \sum \limits_{i=1}^k c \left|B(v_i, r'_i)\right|\\
&\le cn \tag*{since $\{B(v_i, r'_i)\}$ are disjoint}\\
&= O_{\eps}(n). \tag*{\qedhere}
\end{align*}
\end{proof}

We can classify balls as in Lemma \ref{lem:basecluster} into two types as follows:
\begin{lemma} [See Figure \ref{fig:clusters}] \label{lem:clustercat}
For all nodes $v$ and positive integers $r$, we have either:
\begin{itemize}
\item (Large) $|B(v, 2r)| = \Omega\left(r^{4/3} \eee \right)$, or
\item (Bottlenecked) there exists $r < r^* \le 2r$ so that $\left|B_{=}(v, r^*)\right| \le \left|B(v, r^*)\right|^{1/4} \eee^{3/4}$.
\end{itemize}
\end{lemma}
\begin{proof}
Suppose that $B(v, 2r)$ is not bottlenecked.
We have $|B(v, r)| \ge 1$ (conservatively), and we also have the relationship
\begin{align*}
\left|B(v, r^*)\right| &= \left|B(v, r^*-1)\right| + \left|B_{=}(v, r^*)\right|\\
&\ge \left|B(v, r^*-1)\right| + \left|B(v, r^*-1)\right|^{1/4} \eee^{3/4}
\end{align*}
for all $r < r^* \le 2r$ (since $|B_{=}(v, r^*) \ge |B(v_i, r^*-1)|^{1/4} \eee^{3/4}$).
Phrased another way, if we let $x_j := |B(v, r + j)|$, then we have a recurrence with initial condition $x_0 \ge 1$, and recurrence relation
$$x_j \ge x_{j-1} + x_{j-1}^{1/4} \eee^{3/4}.$$
The general solution to a recurrence of this form is
$x_j = \Omega(j^{4/3} \eee)$,
and plugging in $j = r$ thus gives $|B(v, 2r)| = \Omega(r^{4/3} \eee)$, implying that the cluster is large.
%
%
%
\end{proof}

\begin{figure}[ht]
\centering
\begin{minipage}{.4\textwidth}
\begin{tikzpicture}[xscale=0.35, yscale=0.35]
\draw [fill=black, black] (6, 6) circle [radius=0.2];
\node [below=0.1cm] at (6, 6) {$v$};

\draw [thick] (6, 6) -- (10, 6);
\node [above] at (8, 6) {$r$};
\draw [thick] (6, 6) -- ({6+8/1.41421}, {6-8/1.41421});
\node [below=0.1cm] at (6+2/1.41421, 6-2/1.41421) {$2r$};

\foreach \r in {3,...,6}{
	\draw [black] (6, 6) circle [radius=\r*4/3];
}

\foreach \r in {3,...,5}{
	\node [above, black!60!green] at (6, 6+\r*4/3) {BIG};
}

\end{tikzpicture}
\subcaption{In a ``large'' ball, every ring around the core contains many nodes, so the cluster contains many nodes overall.}
\end{minipage}%
\hspace{.1\textwidth}%
\begin{minipage}{.4\textwidth}
\begin{tikzpicture}[xscale=0.35, yscale=0.35]

\draw [thick, red] (6, 6) -- (10.7, 10.7);
\node [red] at (7, 8) {$r^*$};
\draw [fill=black, black] (6, 6) circle [radius=0.2];
\node [below=0.1cm] at (6, 6) {$v$};

\draw [thick] (6, 6) -- (10, 6);
\node [above] at (8, 6) {$r$};
\draw [thick] (6, 6) -- ({6+8/1.41421}, {6-8/1.41421});
\node [below=0.1cm] at (6+2/1.41421, 6-2/1.41421) {$2r$};

\foreach \r in {3,...,6}
	\draw [black] (6, 6) circle [radius=\r*4/3];
	
\node [above, black!60!green] at (6, 6+4) {BIG};
\node [above, red] at (6, 6+4*4/3) {SMALL};
\node [above, black!60!green] at (6, 6+5*4/3) {BIG};
\end{tikzpicture}
\subcaption{In a ``bottlenecked'' ball, at least one of the rings around the core (its ``bottleneck layer'') contains few nodes.  This small layer will be exploited in our construction.}
\end{minipage}
\caption{\label{fig:clusters} The two types of clusters used in our construction of additive spanners.}
\end{figure}

In the following, we let $\{v_1, \dots, v_k\}, \{r_1, \dots, r_k\}$ be selected as in Lemma \ref{lem:basecluster}.
We classify each pair $v_i, r_i$ into \emph{large} or \emph{bottlenecked} as in Lemma \ref{lem:clustercat} (some balls can satisfy both criteria, in which case we can classify as either large or bottlenecked, it doesn't matter).
If it is large, then we refer to the ball $B(v_i, 2r_i)$ as a \emph{large cluster}, and $B(v_i, r_i)$ is its \emph{core}.
If it is bottlenecked, then we refer to the ball $B(v_i, r^*_i)$ as a \emph{bottlenecked cluster}, and $B_{=}(v_i, r^*_i)$ is its \emph{bottleneck layer}.
A node $x \in B(v_i, r^*_i - 1)$ is \emph{bottlenecked} by the cluster $B(v_i, r^*)$.
Note that a node on the bottleneck layer itself is not generally bottlenecked (although it can be bottlenecked if it is also contained in a different cluster).
In either case, $v_i$ is the \emph{center node} of the cluster.

\subsection{Initialization Phase}

Like many spanner constructions, ours can be divided into an \emph{initialization} and \emph{completion} phase.
The completion phase is the part that enforces the final spanner inequality.
The initialization phase adds some edges to the graph beforehand, whose purpose is purely to assist in the argument that the completion phase doesn't add too many edges to the spanner.

In this part we will state the initialization phase (only) and provide some relevant analysis before moving on to the completion phase.
First off, our initialization phase actually includes an additive spanner on the inside.
We will use the following spanners from prior work internally:
\begin{theorem} [\cite{BV16}] \label{thm:prevspanners}
For any $\eee \ge 1$, every $n$-node graph has a spanner on $O(n \eee)$ edges with additive error
$+\Oish\left(n^{1/2} \eee^{-1/2} \right).$
\end{theorem}
The spanners in Theorem \ref{thm:prevspanners} are state-of-the-art only in some range of parameters, when $\eee$ is not too large.
If one uses other spanner constructions instead and rebalances parameters, from \cite{BV16} or \cite{Chechik13soda}, one can optimize the quality of our spanner for certain larger values of $\eee$.
We will not do so, since using piecewise internal spanners introduces considerable additional complexity in the analysis, and the spanners from Theorem \ref{thm:prevspanners} give the best possible results when $\eee = O(1)$ which we consider to be the most interesting setting.
We now state our initialization phase:

\begin{mdframed}[backgroundcolor=gray!10]
\begin{enumerate}
\item \textbf{(Initialization Phase)} The spanner $H$ is initially empty.
Then, we compute a clustering as described in the previous section, and:
\begin{enumerate}
\item For any cluster $C$ of either type, add a shortest path tree rooted at its center node $v$ to all nodes in $C$.

\item For each bottlenecked cluster $B(v, r^*)$, construct a distance preserver with demand pairs $B_{=}(v, r^*)^2$ in the subgraph induced on $B(v, r^*)$, and add its edges to $H$.

\item For each large cluster $B(v, 2r)$, construct a spanner using Theorem \ref{thm:prevspanners} in the subgraph induced on $B(v, 2r)$, and add its edges to $H$.
\end{enumerate}
\item \emph{(Then perform the Completion Phase, described later)}
\end{enumerate}
\end{mdframed}

The distance preservers on bottlenecked clusters are analogous to the step in \cite{BKMP10} in which we add all edges to low-degree nodes.
This roughly lets us ``ignore'' bottlenecked nodes, as we ignore low-degree nodes in \cite{BKMP10}, in a sense we will make precise shortly.
The spanners on large clusters do not have an analogy in \cite{BKMP10}; this is an optimization that is only useful in the setting of large-radius clusters.
The shortest path trees enforce that, if two nodes $x, y$ are in the same cluster $C$ with center node $v$, then we have
$$\dist_H(x, y) \le \dist_H(x, v) + \dist_H(v, y) = \Ohat(\beta);$$
this inequality will occasionally be useful.
Let us next count the number of edges added to the spanner in the initialization phase.

\begin{lemma} [Initialization Edge Bound] \label{lem:initgood}
Assuming Conjecture \ref{cjt:greatddps}, only $O_{\eps}(n \eee)$ edges are added to $H$ during the initialization phase.
\end{lemma}
\begin{proof}
For each bottlenecked cluster $B(v, r^*)$, the associated distance preserver has
$$\left|B_{=}(v, r^*)^2\right| = \left|B_{=}(v, r^*)\right|^2 \le \left| B(v, r^*) \right|^{1/2} \eee^{3/2}$$
demand pairs.
Using Conjecture \ref{cjt:greatddps}, the number of edges in the distance preserver is
$$O\left(|B(v, r^*)|^{2/3} \left(|B(v, r^*)|^{1/2} \eee^{3/2}\right)^{2/3} + |B(v, r^*)| \right) = O(|B(v, r^*)| \eee).$$
For each large cluster $|B(v, 2r)|$, since we apply Theorem \ref{thm:prevspanners} with parameter $\eee$, we add $O(\eee |B(v_i, 2r_i)|)$ edges to the cluster.
We also add a shortest path tree to both types of cluster, but this does not change either estimate.
By a union bound, the total number of edges added over all clusters is thus
\begin{align*}
\sum \limits_{C \text{ is a cluster}} O\left( \eee |C| \right) &= O\left( \eee \cdot \sum \limits_{C \text{ is a cluster}} |C| \right)\\
&= O_{\eps}(n\eee),
\end{align*}
where this last equality follows from the non-overlap property of Lemma \ref{lem:basecluster}.
\end{proof}

Before we state our completion phase, we need a technical lemmas\ describing the structure of shortest paths in the post-initialization spanner.
We think of Lemma \ref{lem:compbot} as a more technically-laden version of the trivial fact used in \cite{BKMP10} that, after we add all edges to low-degree nodes, no missing edges in a shortest path are incident on low-degree nodes; its role in our overall proof is essentially as a tool that lets us focus on large clusters and ignore bottlenecked clusters.
In the following, a \emph{missing edge} is an edge in the original graph $G$ that is not contained in the current spanner (here, meaning after the initialization phase is complete).

\begin{lemma} \label{lem:bottleneckedgood}
For any two non-bottlenecked nodes $s, t$, there is a shortest path $\pi(s, t)$ in $G$ such that no missing edge is contained in a bottlenecked cluster.
\end{lemma}
\begin{proof}
Let $\pi(s, t)$ be an arbitrary shortest path, let $B(v, r^*)$ be a bottlenecked cluster, and suppose $\pi(s, t)$ has a missing edge $e \in B(v, r^*)$.
Let $\pi(s, t)[x \leadsto y]$ be the longest contiguous subpath contained in $B(v, r^*)$ that includes the missing edge $e$.
Since $s, t$ are not bottlenecked, this subpath must terminate in the bottleneck layer; that is, $x, y \in B_{=}(v, r^*)$.
Therefore, we added a shortest $x \leadsto y$ path contained in $B(v, r^*)$ as part of a distance preserver in the initialization phase.
We can thus reroute the subpath $\pi(s, t)[x \leadsto y]$ to coincide with this previously-added shortest $x \leadsto y$ path.
This change at least one missing edge $e$ from $\pi(s, t)$, and it does not introduce any new missing edges to $\pi(s, t)$.
Thus we can repeat this process until the lemma holds.
\end{proof}

\begin{figure} [ht]
\begin{center}
\begin{tikzpicture}
\node [blue] at (-3, 2.5) {$\pi(s, t)$};

\draw [ultra thick] (0, 0) circle [radius=2];
\draw [ultra thick, red] (0, 0) circle [radius=1.5];

\draw [ultra thick, fill=black] (0, 1.75) circle [radius=0.1];
\draw [ultra thick, fill=black] (-1, 1.4) circle [radius=0.1];
\draw [ultra thick, fill=black] (1, 1.4) circle [radius=0.1];

\node [red] at (0, -1.7) {$B_{=}(v, r^*)$};

\draw [ultra thick] plot [smooth] coordinates {(-1, 1.4) (1, 0) (0, 1.75)};
\draw [ultra thick] plot [smooth] coordinates {(1, 1.4) (-1, 0) (0, 1.75)};
\draw [ultra thick] plot [smooth] coordinates {(1, 1.4) (0, -1) (-1, 1.4)};

\draw [ultra thick, dashed, blue, <->] plot [smooth] coordinates {(-3, 2) (-1, 1.4) (0, 0) (1, 1.4) (3, 2)};

\draw [ultra thick, ->] (3, 0) -- (4.5, 0);

\begin{scope}[shift={(7, 0)}]
\node [blue] at (-3, 2.5) {$\pi(s, t)$};

\draw [ultra thick] (0, 0) circle [radius=2];
\draw [ultra thick, red] (0, 0) circle [radius=1.5];

\draw [ultra thick, fill=black] (0, 1.75) circle [radius=0.1];
\draw [ultra thick, fill=black] (-1, 1.4) circle [radius=0.1];
\draw [ultra thick, fill=black] (1, 1.4) circle [radius=0.1];

\node [red] at (0, -1.7) {$B_{=}(v, r^*)$};

\draw [ultra thick] plot [smooth] coordinates {(-1, 1.4) (1, 0) (0, 1.75)};
\draw [ultra thick] plot [smooth] coordinates {(1, 1.4) (-1, 0) (0, 1.75)};
\draw [ultra thick] plot [smooth] coordinates {(1, 1.4) (0, -1) (-1, 1.4)};

\draw [ultra thick, dashed, blue, <->] plot [smooth] coordinates {(-3, 2) (-1, 1.4) (0, -1) (1, 1.4) (3, 2)};

\end{scope}
\end{tikzpicture}
\end{center}

\caption{If a shortest path $\pi(s, t)$ contains a missing edge inside a bottlenecked cluster, then due to the distance preservers added in the initialization phase, we can reroute a subpath of $\pi(s, t)$ to remove this missing edge.  This observation is applied repeatedly to prove Lemma \ref{lem:bottleneckedgood}}
\end{figure}

\begin{lemma} \label{lem:compbot}
For any shortest path $\pi(s, t)$ as in Lemma \ref{lem:bottleneckedgood}, there is a collection of edge-disjoint subpaths $Q$ of $\pi(s, t)$ such that:
\begin{itemize}
\item Every missing edge in $\pi(s, t)$ is contained in a subpath $q \in Q$,

\item Every subpath in $Q$ except possibly for the last one has length $\Omegahat(\beta)$, and

\item Each subpath $q \in Q$ may be ``charged'' to a large cluster $C$ such that $q \subseteq C$, and each large cluster is charged for at most one subpath in $Q$.

\end{itemize}
\end{lemma}
\begin{proof}
Initially $Q \gets \emptyset$.
Repeat the following process until no longer possible: let $(x, y)$ be the first missing edge along $\pi(s, t)$ that is not currently contained in any subpath in $Q$.
By the coverage property of Lemma \ref{lem:basecluster}, we have $x \in B(v, r)$ for some center node $v$ with associated radius $r$.
Let $q$ be the subpath of $\pi(s, t)$ that starts with the edge $(x, y)$ and extends as far as possible under the constraint $q \subseteq B(v, 2r)$.
Add $q$ to $Q$, then repeat until all missing edges along $\pi(s, t)$ are covered by subpaths in $Q$.

It is immediate from the construction that our subpaths are edge-disjoint and that they cover the missing edges of $\pi(s, t)$.
We next verify the subpath length property: a selected subpath $q$ starts at a node $x \in B(v, r)$, and it ends when it cannot be extended further, either because it reaches the end of $\pi(s, t)$ (in which case it is the last subpath) or because it reaches $B_{=}(v, 2r)$ (in which case it has length at least $r = \Omegahat(\beta)$).

Next, we verify the subpath charging property.
For each missing edge $(x, y)$, we have $x \in B(v, r)$.
Notice that $v$ cannot be the center of a \emph{bottlenecked} cluster, since otherwise the missing edge $(x, y)$ would be contained in this bottlenecked cluster, but by Lemma \ref{lem:bottleneckedgood} no missing edge in $\pi(s, t)$ is in a bottlenecked cluster.
Thus $v$ is the center node of a \emph{large} cluster, and we can charge the associated subpath $q$ to this large cluster, since we have $q \subseteq B(v, 2r)$.

Finally, we verify that each large cluster is only charged for one subpath in $Q$.
Suppose for contradiction that a large cluster $B(v, 2r)$ is charged for two different subpaths $q_1, q_2$, due to two different missing edges $(x_1, y_1), (x_2, y_2)$ with $x_1, x_2 \in B(v, r)$.
That means there is a node $z \in \pi(s, t)[x_1 \leadsto y_2]$ with $z \notin B(v, 2r)$, since otherwise the subpath $q_1$ would extend until it covers $(x_2, y_2)$ as well.
So we have
$$\dist_G(x_1, x_2) = \dist_G(x_1, z) + \dist_G(z, x_2) > 2r,$$
since $x_1, x_2 \in B(v, r)$ but $z \notin B(v, 2r)$.
However, by the triangle inequality, we also have
$$\dist_G(x_1, x_2) \le \dist_G(x_1, v) + \dist_G(v, x_2) \le 2r,$$
completing the contradiction.
\end{proof}

\subsection{Completion Phase}

We now state the completion phase of the algorithm:

\begin{mdframed} [backgroundcolor=gray!10]
\begin{enumerate}
\item \emph{(Perform the Initialization Phase, described previously)}\\

\hspace{-1cm} \textbf{(Completion Phase:)}

\item  Let $c$ be a sufficiently large positive constant.
For each pair of non-bottlenecked nodes $s, t$, if currently
$$\dist_H(s, t) > \dist_G(s, t) + \beta n^{c \eps}:$$
\begin{enumerate}
\item Let $\pi(s, t), Q$ be a shortest path and collection of subpaths as in Lemma \ref{lem:compbot}.

\item For each subpath $q \in Q$, letting $x, y$ be its endpoints, add $(x, y)$ as a \textbf{weighted} edge to the spanner $H$, with weight $\dist_G(x, y)$.
Charge $(x, y)$ to the large cluster to which we charged $q$.
\end{enumerate}

\item For each large cluster $C$, let $P_C$ be the set of weighted edges charged to $C$.
Replace the edges $P_C$ with a distance preserver on the subgraph induced on $C$, with demand pairs $P_C$.

\item Return the spanner $H$.
\end{enumerate}
\end{mdframed}

One can imagine our completion phase as following the greedy completion strategy from \cite{BKMP10} or its update in \cite{Knudsen14}: while there is a pair of non-bottlenecked nodes $s, t$ that violate the spanner inequality, add a shortest path $\pi(s, t)$ to the spanner to correct this.
But there are two key differences.
First, while \cite{BKMP10} uses a ``path-buying'' argument to directly bound the number of edges added to each cluster, we use path-buying to bound the \emph{number of subpaths} that get ``charged'' to each large cluster.
We will then apply distance preserver bounds at the end to convert a bound on number of subpaths per cluster to a bound on number of edges per cluster.

Second, in order to apply these distance preserver bounds, we need to be very careful about our choice of shortest subpaths through each large cluster.
That motivates the use of our intermediate weighted edges: these act as ``placeholders'' for the specific shortest path through a large cluster that will ultimately be selected; they record that we have committed to enforcing $\dist_H(x, y) = \dist_G(x, y)$ in the final spanner, even though we haven't yet picked the shortest path that will achieve this.
We resolve our weighted edges into shortest paths at the end: this is a way to delay our choice of shortest $x \leadsto y$ path until we know the \emph{entire} set of shortest paths in a large cluster that will need to be selected, which lets us use a distance preserver construction.
If we tried to simplify our algorithm by avoiding weighted edges, instead having it commit to an entire shortest path $\pi(s, t)$ right when the node pair $s, t$ is considered, then we would only be able to use ``online'' distance preserver bounds, where the shortest path for each demand pair is selected before the next demand pair arrives.

We will now begin to formally prove properties of our construction.
We begin with correctness of the output spanner:

\begin{lemma} [Spanner Correctness] \label{lem:spcorrect}
For all nodes $s, t$, we have
$\dist_H(s, t) \le \dist_G(s, t) + \Ohat \left( \beta \right).$
\end{lemma}
\begin{proof}
Let $\pi(s, t)$ be a shortest $s \leadsto t$ path as in Lemma \ref{lem:bottleneckedgood}.
Let $s'$ be the last node on $\pi(s, t)$ such that there exists a bottlenecked cluster $C$ with $s, s' \in C$, or if $s$ is not contained in any bottlenecked cluster then let $s' := s$.
In either case, since we added a shortest path tree rooted at the center node of $C$, the distance between $s, s'$ is at most twice the radius of $C$; that is,
$$\dist_H(s, s') = \Ohat(\beta).$$
Similarly, we let $t'$ be the first node on $\pi(s, t)$ such that there exists a bottlenecked cluster $C$ with $t, t' \in C$, or if $t$ is not contained in any bottlenecked cluster then let $t' := t$.
For the same reason we have
$$\dist_H(t, t') = \Ohat(\beta).$$
Next, let $s'', t''$ be the first, last non-bottlenecked nodes along the subpath $\pi(s, t)[s' \leadsto t']$.
Notice that all edges on the $s' \leadsto s''$ and $t' \leadsto t''$ subpaths are contained in a bottlenecked cluster, but by choice of $s', t'$ these bottlenecked clusters do not also contain $s, t$.
Thus, by Lemma \ref{lem:bottleneckedgood}, every edge along the $s' \leadsto s''$ and $t' \leadsto t''$ subpaths is already contained in the spanner, and so we have
$$\dist_H(s', s'') = \dist_G(s', s'') \text{ and } \dist_H(t', t'') = \dist_G(t', t'').$$
Finally, in the completion phase, since $s'', t''$ are non-bottlenecked we ensure that
$$\dist_H(s'', t'') \le \dist_G(s'', t'') + \Ohat\left( \beta \right).$$
Putting it all together, we have:
\begin{align*}
\dist_H(s, t) &\le \dist_H(s, s') + \dist_H(s', s'') + \dist_H(s'', t'') + \dist_H(t'', t') + \dist_H(t', t)\\
&= \dist_H(s', s'') + \dist_H(s'', t'') + \dist_H(t'', t') + \Ohat\left( \beta \right)\\
&= \dist_G(s', s'') + \dist_H(s'', t'') + \dist_G(t'', t') + \Ohat\left( \beta \right)\\
&= \dist_G(s', s'') + \dist_G(s'', t'') + \dist_G(t'', t') + \Ohat\left( \beta \right)\\
&= \dist_G(s', t') + \Ohat\left( \beta \right)\\
&\le \dist_G(s, t) + \Ohat\left( \beta \right). \tag*{\qedhere}
\end{align*}
\end{proof}

\begin{figure}
\begin{center}
\begin{tikzpicture}
\draw [ultra thick] (-5, 0) circle [radius=2];
\draw [ultra thick] (5, 0) circle [radius=2];

\draw [fill=black] (-5, 0) circle [radius=0.15];
\draw [fill=black] (5, 0) circle [radius=0.15];
\node at (-5, -1) {(bottlenecked)};
\node at (5, -1) {(bottlenecked)};

\node at (-6, 1.5) {$s$};
\draw [fill=blue] (-6, 1) circle [radius=0.15];

\node at (6, 1.5) {$t$};
\draw [fill=blue] (6, 1) circle [radius=0.15];

\draw [fill=blue] (-3.3, 1) circle [radius=0.15];
\node at (-3.3, 1.5) {$s'$};

\draw [fill=blue] (3.3, 1) circle [radius=0.15];
\node at (3.3, 1.5) {$t'$};

\draw [fill=blue] (-2, 1) circle [radius=0.15];
\node at (-2, 1.5) {$s''$};

\draw [fill=blue] (2, 1) circle [radius=0.15];
\node at (2, 1.5) {$t''$};

\draw [ultra thick, blue] (-6, 1) -- (-5, 0) -- (-3.3, 1) -- (3.3, 1) -- (5, 0) -- (6, 1);

\node at (-6, 0.5) {$\Ohat(\beta)$};
\node at (6, 0.4) {$\Ohat(\beta)$};
\node at (-3.5, 0.4) {$\Ohat(\beta)$};
\node at (3.6, 0.4) {$\Ohat(\beta)$};

\node at (0, 1.4) {$+\Ohat(\beta)$ err};
\node at (0, 0.6) {(completion phase)};

\draw [thick] (0, -2) -- (-2.6, 0.7);
\draw [thick] (0, -2) -- (2.6, 0.7);
\node [align=center, fill=white] at (0, -2) {$+0$ err\\ all edges in spanner \\ (Lemma \ref{lem:bottleneckedgood})};

\end{tikzpicture}
\end{center}
\caption{The organization of the subpaths analyzed in our proof of spanner correctness (Lemma \ref{lem:spcorrect}).}
\end{figure}

Our goal is now to control the number of edges added in the completion phase.
As discussed previously, we generally follow the path-buying method from \cite{BKMP10}, with one new optimization for our setting.
In \cite{BKMP10}, the strategy is roughly to argue that each time a shortest path $\pi(s, t)$ is added, for each cluster $C$ that is hit by $\pi(s, t)$, we improve the spanner distance from $C$ to either the first or the last cluster hit by $\pi(s, t)$.
This is used to control the number of times any given cluster is hit by an added shortest path.
Our strategy is similar, but we can gain a bit by considering not just the first and last large clusters hit by $\pi(s, t)$, but rather the \emph{first few} and \emph{last few} large clusters hit by $\pi(s, t)$.
This finally reveals the purpose of the spanners we added to large clusters in the initialization phase: essentially, they allow us to traverse a few extra clusters at the start and end and stay within our error budget.

Let us next formalize what we mean by ``first few'' and ``last few'' large clusters.
For a shortest path $\pi(s, t)$ with associated subpath list $Q$ selected in the completion phase, we define the \emph{prefix} $Q_{\text{pre}} \subseteq Q$ to be the minimal prefix of subpaths in $Q$ that satisfies
$$\sum \limits_{\substack{C \text{ large cluster}\\ \text{some } q \in Q_{\text{pre}} \text{ charged to } C}} |C| \ge \beta^{5/3} \eee.$$
To avoid confusion around the word ``prefix:'' note that $Q_{\text{pre}}$ is \textbf{not} obtained by selecting some prefix of each subpath in $Q$; rather, each subpath in $Q$ is entirely in or entirely out of $Q_{\text{pre}}$.
We are taking the first few subpaths in $Q$ used by the path being analyzed, until they have sufficiently large total size.
Similarly, we define the \emph{suffix} $Q_{\text{suf}}$ as the minimal suffix of subpaths in $Q$ that satisfies
$$\sum \limits_{\substack{C \text{ large cluster}\\ \text{some } q \in Q_{\text{suf}} \text{ charged to } C}} |C| \ge \beta^{5/3} \eee,$$
and we again caution that this is \textbf{not} obtained by taking a suffix of each path in $Q$, but rather, taking all of the last few subpaths in the ordering of $Q$.

If no prefix/suffix of $Q$ achieves this inequality, then we can define $Q_{\text{pre}} = Q_{\text{suf}} = Q$, but it turns out we will not encounter this situation in our algorithm so it won't be too important (Lemma \ref{lem:presuferr} to follow implies that, if the prefix and suffix overlap, then the node pair $s, t$ is already well-spanned and so it won't be selected in the completion phase).
It is technically possible for a node to be contained in several large clusters charged by prefix/suffix subpaths.
However, we can bound the frequency with which this happens, leading to a bound on the number of \emph{distinct} nodes in these prefix/suffixes.
Let $A_{\text{pre}}$ be the set of nodes contained in any large cluster $C$ for which some $q \in Q_{\text{pre}}$ is charged to $C$, and similarly let $A_{\text{suf}}$ be the set of nodes contained in any large cluster $C$ for which some $q \in Q_{\text{suf}}$ is charged to $C$.

\begin{lemma} [Prefix/Suffix Size] \label{lem:shpnbhd}
$|A_{\text{pre}}|, |A_{\text{suf}}| = \Omegahat \left( \beta^{5/3} \eee \right)$.
\end{lemma}
\begin{proof}
Let $v$ be a node in $A_{\text{pre}}$ or $A_{\text{suf}}$.
Let $x, y$ be the first, last nodes on $\pi(s, t)$ that are respectively contained in large clusters $C_x, C_y$ with $v \in C_x, v \in C_y$.
Every large cluster has radius $\Ohat(\beta )$, and so we have
$$\dist_G(x, v), \dist_G(y, v) = \Ohat \left(\beta \right).$$
So by the triangle inequality, we have
$\dist_G(x, y) = \Ohat \left( \beta \right).$
From Lemma \ref{lem:compbot}, every subpath in $Q$ has length $\Omegahat(\beta)$, except possibly for the last one.
Hence there can only be $\Ohat(1)$ total subpaths in $Q$ that contain nodes between $x$ and $y$, and so we can charge only $\Ohat(1)$ different large clusters that contain $v$.
Thus $v$ is counted only $\Ohat(1)$ times in the sum used to define the prefix and suffix, leading to the claimed bound.
\end{proof}

When we say the prefix/suffix can be traversed within our error budget, we mean the following lemma:
\begin{lemma} [Prefix/Suffix Distance Error] \label{lem:presuferr}
For any $x \in A_{\text{pre}}, y \in A_{\text{suf}}$, we have
$$\dist_H(s, x) \le \dist_G(s, x) + \Ohat \left(\beta \right) \text{ and } \dist_H(y, t) \le \dist_G(y, t) + \Ohat \left(\beta \right).$$
\end{lemma}
\begin{proof}
We will prove the inequality for the prefix here ($x \in A_{\text{pre}}$); the inequality for the suffix is essentially identical.
Let $C^*$ be the large cluster charged by a subpath $q^* \in Q_{\text{pre}}$ with $x \in C^*$, and let $x' \in \pi(s, t)$ be the first node on the subpath $q^*$.
Let $\{q_1, \dots, q_k, q^*\}$ be the leading subpaths of $Q_{\text{pre}}$ up to $q^*$, and let $\{C_1, \dots, C_k, C^*\}$ be the respective large clusters charged for these subpaths.
Let $x_i, y_i$ be the endpoints of each subpath $q_i$.
We then have
\begin{align*}
\sum \limits_{i=1}^{k-1} \dist_H(x_i, y_i) &\le \sum \limits_{i=1}^{k-1} \dist_G(x_i, y_i) + \Oish\left(\frac{|C_i|^{1/2}}{\eee^{1/2}}\right) \tag*{Spanners from initialization}\\
&\le \sum \limits_{i=1}^{k-1} \dist_G(x_i, y_i) + \Oish\left(\frac{|C_i|}{\beta^{2/3} \eee}\right) \tag*{since $|C_i| = \Omega(\beta^{4/3} \eee)$ by Lemma \ref{lem:clustercat}}\\
&\le \left( \sum \limits_{i=1}^{k-1} \dist_G(x_i, y_i) \right) + \Oish\left(\frac{\beta^{5/3} \eee}{\beta^{2/3} \eee}\right) \tag*{Def of prefix}\\
&\le \left( \sum \limits_{i=1}^{k-1} \dist_G(x_i, y_i) \right) + \Oish\left(\beta \right).
\end{align*}
From Lemma \ref{lem:compbot}, every edge along $\pi(s, t)[s \leadsto x]$ except those contained in $x_i \leadsto y_i$ subpaths is already contained in the spanner.
So we have
\begin{align*}
\dist_H(s, x') &\le \dist_H(s, x_1) + \left(\sum \limits_{i=1}^{k} \dist_H(x_i, y_i) \right) + \left(\sum \limits_{i=1}^{k-1} \dist_H(y_i, x_{i+1}) \right) + \dist_H(y_k, x')\\
&= \dist_G(s, x_1) + \left(\sum \limits_{i=1}^{k} \dist_G(x_i, y_i) + \Oish \left(\beta \right)\right) + \left(\sum \limits_{i=1}^{k-1} \dist_G(y_i, x_{i+1}) \right) + \dist_G(y_k, x')\\
&= \dist_G(s, x') + \Oish\left( \beta \right).
\end{align*}
Finally, we can complete the proof using the triangle inequality:
\begin{align*}
\dist_H(s, x) &\le \dist_H(s, x') + \dist_H(x', x)\\
&\le \left(\dist_G(s, x') + \Oish(\beta)\right) + \dist_H(x', x)\\
&\le \dist_G(s, x) + \dist_G(x, x') + \dist_H(x', x) + \Oish(\beta)\\
&\le \dist_G(s, x) + \Ohat(\beta) \tag*{$x, x'$ in same large cluster. \qedhere}
\end{align*}
\end{proof}

In our next two lemmas, we will focus on a particular large cluster $L$ that is charged by at least one subpath $q_L \in Q$.
We say that $L$ is \emph{cluster-preserved} with a node $x$ if there exists a node $\ell \in L$ and a node $x'$ such that $x, x'$ are in the same large cluster and $\dist_H(\ell, x') = \dist_G(\ell, x')$.
We use cluster-preservation as a sort of potential function: we will argue that, when we charge a new weighted edge to $L$, we also cause $L$ to become cluster-preserved with many new nodes.
This limits the total number of weighted edges that may be charged to $L$ over the entire completion phase, since $L$ can only become newly cluster-preserved with $n$ nodes in total.

\begin{lemma} \label{lem:precp}
Before the node pair $s, t$ is considered, either $L$ is not cluster-preserved with any node in $A_{\text{pre}}$, or $L$ is not cluster-preserved with any node in $A_{\text{suf}}$.
\end{lemma}
\begin{proof}
Let us first handle the case where the subpath $q_L$ charging $L$ is in neither the prefix nor the suffix.
Suppose for contradiction that there are nodes $x \in A_{\text{pre}}, y \in A_{\text{suf}}$ and $\ell_x, \ell_y \in L$ with
$$\dist_H(x, \ell_x) = \dist_G(x, \ell_x) \text{ and } \dist_H(y, \ell_y) = \dist_G(y, \ell_y).$$
Let $x', y'$ be nodes on $\pi(s, t)$ that share a large cluster with $x, y$ respectively, and let $\ell \in q_L$.
We then have
\begin{align*}
\dist_H(s, t) &\le \dist_H(s, x) + \dist_H(x, \ell_x) + \dist_H(\ell_x, \ell_y) + \dist_H(\ell_y, y) + \dist_H(y, t)\\
&\le \dist_H(s, x) + \dist_G(x, \ell_x) + \dist_G(\ell_y, y) + \dist_H(y, t) + \Ohat \left(\beta \right) \tag*{$\ell_x, \ell_y$ in same cluster}\\
&\le \dist_G(s, x) + \dist_G(x, \ell_x) + \dist_G(\ell_y, y) + \dist_G(y, t) + \Ohat \left(\beta \right) \tag*{Lemma \ref{lem:presuferr}}\\
&\le \dist_G(s, x') + \dist_G(x', \ell) + \dist_G(\ell, y') + \dist_G(y', t) + \Ohat \left(\beta \right)
\end{align*}
where the last inequality uses the triangle inequality together with the fact that the nodes $(x, x'), (\ell_x, \ell, \ell_y), (y, y')$ each lie in same large cluster, which has radius $\Ohat(\beta )$.
Since $x', \ell, y'$ all lie along $\pi(s, t)$, this implies
$$\dist_H(s, t) \le \dist_G(s, t) + \Ohat \left( \beta \right).$$
But, by choice of large enough constant $c$, this means we wouldn't select the node pair $s, t$ in the completion phase and add any corresponding weighted edges, since the distance inequality for this node pair would be already satisfied.
This contradicts the assumption that $s, t$ were selected in the completion phase.

The case where $q_L$ is in the prefix follows an essentially identical analysis, which we will not repeat: in this case we do not need nodes $x, x'$, and we specifically argue that $L$ is not cluster-preserved with any node in $A_{\text{suf}}$ (or else the above inequality holds, again contradicting that $s, t$ were selected).
In the case where $q_L$ is in the suffix, we instead do not need nodes $y, y'$, and we specifically argue that $L$ is not cluster-preserved with any node in $A_{\text{pre}}$.
\end{proof}

\begin{lemma} \label{lem:postcp}
After the node pair $s, t$ is considered, $L$ is cluster-preserved with every node in $A_{\text{pre}} \cup A_{\text{suf}}$.
\end{lemma}
\begin{proof}
Let $\pi(s, t)$ be the shortest path in $G$, and let $\pi^*(s, t)$ be the shortest path in the spanner $H$ obtained from $\pi(s, t)$ by replacing each $x \leadsto y$ subpath in $Q$ with the weighted edge $(x, y)$ added to the spanner.
(Notice that, from Lemma \ref{lem:compbot}, every edge in $\pi(s, t)$ not contained in a subpath in $Q$ is already in $H$ from the initialization phase, so this does indeed describe a path contained in $H$.)

Let $z \in A_{\text{pre}} \cup A_{\text{suf}}$, and let $C$ be the large cluster containing $z$ that is charged by a prefix or suffix subpath.
By construction there is a weighted edge $(x, y) \in \pi^*(s, t)$ with $x, y \in C$.
Similarly, there is a weighted edge $(x', y') \in \pi^*(s, t)$ with $x', y' \in L$.
Thus we have $\dist_H(x, x') = \dist_G(x, x')$, due to the shortest path $\pi^*(s, t)$.
So $L$ is cluster-preserved with $z$ after $\pi^*(s, t)$ is added.
\end{proof}

We now put the pieces together:
\begin{lemma} \label{lem:compgood}
Assuming Conjecture \ref{cjt:greatddps}, the number of edges added to $H$ during the completion phase is
$$O_{\eps}(n) + \Ohat\left(\frac{n^{5/3}}{\beta^{14/9} \eee} \right).$$
\end{lemma}
\begin{proof}
Each time we charge a weighted edge to a large cluster $L$, by Lemmas \ref{lem:precp} and \ref{lem:postcp}, we also newly cluster-preserve it with all nodes in $A_{\text{pre}}$ or with all nodes in $A_{\text{suf}}$.
Using Lemma \ref{lem:shpnbhd}, we have
$$|A_{\text{suf}}|, |A_{\text{pre}}| = \Omegahat \left( \beta^{5/3} \eee \right).$$
Thus, the total number of weighted edges charged to $L$ is only
$$\Ohat \left( \frac{n}{\beta^{5/3} \eee} \right).$$
Applying Conjecture \ref{cjt:greatddps}, the number of edges added to $L$ as part of the final distance preserver is
\begin{align*}
&O\left( |L| + |L|^{2/3} \Ohat\left(\frac{n}{\beta^{5/3} \eee} \right)^{2/3} \right)\\
=& O\left( |L| + |L| \left( \beta^{4/3} \eee \right)^{-1/3} \Ohat\left(\frac{n^{2/3}}{\beta^{10/9} \eee^{2/3}} \right) \right) \tag*{since $|L| = \Omega(\beta^{4/3} \eee)$ by Lemma \ref{lem:clustercat}}\\
=& O\left(|L|\right) + \Ohat \left(|L| \cdot \frac{n^{2/3}}{\beta^{14/9} \eee} \right).
\end{align*}
Hence the total number of edges added in the completion phase is
\begin{align*}
&\sum \limits_{L \text{ large cluster}} O\left(|L|\right) + \Ohat \left(|L| \cdot \frac{n^{2/3}}{\beta^{14/9} \eee} \right)\\
=& O_{\eps}(n) + \Ohat\left(\frac{n^{5/3}}{\beta^{14/9} \eee} \right) \tag*{Non-overlap property of Lemma \ref{lem:basecluster}. \qedhere}
\end{align*}
\end{proof}

\begin{figure}[ht]
\centering
\begin{tikzpicture}
\draw [fill=black] (0, 0) circle [radius=0.15cm];
\node at (-0.5, 0) {$s$};
\draw [fill=black] (12, 0) circle [radius=0.15cm];
\node at (12.5, 0) {$t$};
\draw [fill=black] (1, 0) circle [radius=0.15cm];
\draw [fill=black] (2, 0) circle [radius=0.15cm];
\draw [fill=black] (3, 0) circle [radius=0.15cm];
\draw [ultra thick, dotted] (0, 0) -- (1, 0);
\draw [ultra thick] (1, 0) -- (2, 0);
\draw [ultra thick, dotted] (2, 0) -- (3, 0);

\draw [fill=black] (9, 0) circle [radius=0.15cm];
\draw [fill=black] (10, 0) circle [radius=0.15cm];
\draw [fill=black] (11, 0) circle [radius=0.15cm];

\draw [ultra thick, dotted] (9, 0) -- (10, 0);
\draw [ultra thick] (10, 0) -- (11, 0);
\draw [ultra thick, dotted] (11, 0) -- (12, 0);


\draw [thick, blue] (1,0) ellipse (2 and 1.5);
\node [blue] at (1, 2) {$|A_{\text{pre}}| = \Omegahat\left(\beta^{5/3} \eee\right)$};
\draw [thick, blue] (11,0) ellipse (2 and 1.5);
\node [blue] at (11, 2) {$|A_{\text{suf}}| = \Omegahat\left(\beta^{5/3} \eee\right)$};

\draw [fill=black] (5.5, 1.5) circle [radius=0.15cm];
\node at (5.5, 2) {$\ell_x$};
\draw [fill=black] (6.5, 1) circle [radius=0.15cm];
\node at (6.5, 1.5) {$\ell_y$};

\draw [fill=black] (2, 0.5) circle [radius=0.15cm];
\node at (1.6, 0.5) {$x$};
\draw [fill=black] (10, 0.5) circle [radius=0.15cm];
\node at (10.4, 0.5) {$y$};

\draw [thick, dashed] (5.5, 1.5) -- (2, 0.5);
\draw [thick, dashed] (6.5, 1) -- (10, 0.5);

\node [red] at (4.5, 1.2) {\Huge \bf $\times$};
\node [red] at (7.5, 0.9) {\Huge \bf $\times$};
\draw [red] (5.5, 0.4) -- (4.5, 1.2);
\draw [red] (6.5, 0.4) -- (7.5, 0.9);

\node [red, fill=white] at (6, 0.4) {not both};

\draw [ultra thick] (6, 0.9) circle [radius=1.5];
\node at (6, 3) {$L$};

\node [rotate=15] at (3.5, 1.2) {\small cluster-preserved};
\node [rotate=-10] at (8.3, 1) {\small cluster-preserved};

\draw [ultra thick, fill=black] (4.8, 0) circle [radius=0.15cm];
\draw [ultra thick, fill=black] (7.2, 0) circle [radius=0.15cm];
\draw [ultra thick] (3, 0) -- (4.8, 0);
\draw [ultra thick] (7.2, 0) -- (9, 0);
\draw [ultra thick, dashed, blue] (4.8, 0) -- (7.2, 0);

\node [blue] at (6, -0.3) {\small charged};

\end{tikzpicture}

\caption{\label{fig:connect} The ``path-buying'' part of our argument argues that each time we charge a new weighted edge to a large cluster $L$, we also cluster-preserve $L$ with many new nodes (either $A_{\text{pre}}$ or $A_{\text{suf}}$) for the first time, thus limiting the total number of weighted edges charged to $L$.}
\end{figure}

We now finish the proof with a parameter balance.
Setting
$$\beta = \Thetahat\left(n^{3/7} \eee^{-9/7}\right)$$
means the number of edges added in the completion phase is
\begin{align*}
&O_{\eps}(n) + O\left(\frac{n^{5/3}}{\left( n^{3/7} \eee^{-9/7} \right)^{14/9} \eee} \right)\\
=& O_{\eps}\left( n + \frac{n^{5/3}}{n^{2/3} \eee^{-2} \eee} \right)\\
&= O_{\eps}\left(n \eee \right),
\end{align*}
which completes the proof of Theorem \ref{thm:adddspan}.

\subsection{Unconditional Extensions \label{sec:uncond}}

Theorem \ref{thm:adddspan} is phrased conditionally on Conjecture \ref{cjt:greatddps}.
Here we discuss unconditional extensions of our result, that rely only on known distance preserver bounds.
In order to work generally with all possible distance preserver bounds at once, in the following we let $a, b$ be absolute constants such that for any $n$-node graph and set of $p$ demand pairs, there is a distance preserver on $O(n + n^a p^b)$ edges.
Thus \cite{CE06} shows that we may take $a=1/2, b=1$ or $a=1, b=1/2$, and under Conjecture \ref{cjt:greatddps} we may take $a=b=2/3$.
We then change the following parts of the preceding argument:

\begin{enumerate}
\item Distance preserver bounds influence Lemma \ref{lem:clustercat}, where we classify clusters into large or bottlenecked.
The specific choice of parameters used for bottlenecked clusters comes from our need to argue in Lemma \ref{lem:initgood} that each bottlenecked cluster $C$ costs only $O(|C| \eee)$ edges for its distance preserver in the initialization.
So, we need to redefine bottlenecked clusters in such a way that this still holds.
Specifically, we say that a cluster is bottlenecked if there exists $r < r^* \le 2r$ so that
$$\left| B_{=}(v, r^*)\right| \le \left| B(v, r^*) \right|^{\frac{1-a}{2b}} \eee^{\frac{1}{2b}}.$$
That way, the cost of the distance preserver for a bottlenecked cluster is
\begin{align*}
& O\left( \left| B(v, r^*) \right| + \left| B(v, r^*) \right|^a \left|B_{=}(v, r^*)\right|^{2b}\right)\\
=& O\left( \left| B(v, r^*) \right| + \left| B(v, r^*) \right|^a \left(\left| B(v, r^*) \right|^{\frac{1-a}{2b}} \eee^{\frac{1}{2b}}\right)^{2b}\right)\\
=& O\left( \left| B(v, r^*) \right| + \left| B(v, r^*) \right|^a \left| B(v, r^*) \right|^{1-a} \eee \right)\\
=& O\left( \left| B(v, r^*) \right| \eee \right),
\end{align*}
as desired.

\item This redefinition of bottlenecked clusters influences the associated size bound for large clusters.
Revisiting the proof of Lemma \ref{lem:clustercat}, if $B(v, 2r)$ is not bottlenecked, then we can again describe its size using a recurrence where $x_j := \left|B(v, r+j)\right|$.
The initial condition is $x_0 \ge 1$, and the recurrence relation is
$$x_j \ge x_{j-1} + x_{j-1}^{\frac{1-a}{2b}} \eee^{\frac{1}{2b}}.$$
This recurrence solves asymptotically to
$$x_r = \Omega\left( r^{\frac{2b}{2b+a-1}} \eee^{\frac{1}{2b+a-1}} \right).$$
So this is our new size lower bound for large clusters.

\item The definition of the prefix and suffix $Q_{\text{pre}}, Q_{\text{suf}}$ use a particular bound on the sum of cluster sizes.
This bound is selected to push through Lemma \ref{lem:presuferr}, which lets us reach prefix and suffix nodes within small error.
This is a function of both the spanner quality in Theorem \ref{thm:prevspanners} and the sizes of large clusters.
Our new definition of the prefix $Q_{\text{pre}} \subseteq Q$ is the minimal prefix of subpaths satisfying
$$\sum \limits_{\substack{C \text{ large cluster}\\ \text{some } q \in Q_{\text{pre}} \text{ charged to } C}} |C| \ge \beta^{\frac{3b+a-1}{2b+a-1}} \eee^{\frac{2b+a}{4b+2a-2}}$$
and the suffix $Q_{\text{suf}} \subseteq Q$ is defined similarly.
To re-prove Lemma \ref{lem:presuferr} under this new definition of the prefix and suffix, we proceed through the proof exactly as before, and recompute the following intermediate step:
\begin{align*}
\sum \limits_{i=1}^{k-1} \dist_H(x_i, y_i) &\le \sum \limits_{i=1}^{k-1} \dist_G(x_i, y_i) + \Oish\left(|C_i|^{1/2} \eee^{-1/2}\right) \tag*{Spanners from initialization}\\
&\le \sum \limits_{i=1}^{k-1} \dist_G(x_i, y_i) + \Oish\left(|C_i| \left(\beta^{\frac{2b}{2b+a-1}} \eee^{\frac{1}{2b+a-1}}\right)^{-1/2}  \eee^{-1/2} \right) \tag*{Large cl.\ size}\\
&\le \sum \limits_{i=1}^{k-1} \dist_G(x_i, y_i) + \Oish\left(|C_i| \cdot \beta^{\frac{-b}{2b+a-1}} \eee^{\frac{-2b-a}{4b+2a-2}} \right) \\
&\le \left(\sum \limits_{i=1}^{k-1} \dist_G(x_i, y_i)\right) + \Oish\left(\left(\beta^{\frac{3b+a-1}{2b+a-1}} \eee^{\frac{2b+a}{4b+2a-2}} \right) \beta^{\frac{-b}{2b+a-1}} \eee^{\frac{-2b-a}{4b+2a-2}} \right) \tag*{Def of prfx} \\
&\le \left(\sum \limits_{i=1}^{k-1} \dist_G(x_i, y_i)\right) + \Oish\left(\beta \right). \\
\end{align*}
From there, the rest of the proof is the same as before.

\item Lemma \ref{lem:compgood} uses distance preserver bounds to convert our limit on the number of weighted edges charged to each large cluster to a bound on the number of edges in its associated distance preserver.
Specifically, we get
$$|A_{\text{suf}}|, |A_{\text{pre}}| = \Omegahat\left( \beta^{\frac{3b+a-1}{2b+a-1}} \eee^{\frac{2b+a}{4b+2a-2}} \right),$$
and so the number of weighted edges charged to any given large cluster $L$ is
$$\Ohat\left(\frac{n}{\beta^{\frac{3b+a-1}{2b+a-1}} \eee^{\frac{2b+a}{4b+2a-2}}} \right).$$
Applying our distance preserver bound of $O(n + n^a p^b)$, the number of edges added to $L$ as part of the final distance preserver is
\begin{align*}
&O\left(|L| + |L|^a \cdot \Ohat\left( \frac{n}{\beta^{\frac{3b+a-1}{2b+a-1}} \eee^{\frac{2b+a}{4b+2a-2}}} \right)^b \right)\\
=& O\left(|L| + |L| \cdot \left( \beta^{\frac{2b}{2b+a-1}} \eee^{\frac{1}{2b+a-1}} \right)^{a-1} \cdot \Ohat\left( \frac{n}{\beta^{\frac{3b+a-1}{2b+a-1}} \eee^{\frac{2b+a}{4b+2a-2}}} \right)^b \right) \tag*{Large cl.\ size}\\
=& O\left(|L|\right) + \Ohat\left( |L| \cdot \left( \beta^{\frac{2ab - 2b}{2b+a-1}} \eee^{\frac{a-1}{2b+a-1}} \right) \cdot \left(\frac{n^b}{\beta^{\frac{3b^2 + ab - b}{2b+a-1}} \eee^{\frac{2b^2 + ab}{4b+2a-2}}} \right) \right) \\
=& O\left(|L|\right) + \Ohat\left( |L| \cdot n^b \cdot \beta^{\frac{-3b^2 + ab - b}{2b+a-1}} \eee^{\frac{-2b^2 - ab + 2a - 2}{4b+2a-2}} \right).
\end{align*}
Summing over all large clusters and applying the non-overlap property as before, we get that the total number of edges added in the completion phase is
$$O_{\eps}\left(n\right) + \Ohat\left( n^{1+b} \cdot \beta^{\frac{-3b^2 + ab - b}{2b+a-1}} \eee^{\frac{-2b^2 - ab + 2a - 2}{4b+2a-2}} \right).$$

\item Finally, we need to revisit our setting of $\beta$ in the final parameter balance.
We set
$$\beta = \Ohat\left( n^{\frac{2b + a - 1}{3b - a + 1}} \eee^{\frac{-2b-a-4}{6b-2a+2}} \right),$$
and so the number of edges added in the completion phase is
\begin{align*}
&O_{\eps}\left(n\right) + \Ohat\left( n^{1+b} \cdot \left( n^{\frac{2b + a - 1}{3b - a + 1}} \eee^{\frac{-2b-a-4}{6b-2a+2}} \right)^{\frac{-3b^2 + ab - b}{2b+a-1}} \eee^{\frac{-2b^2 - ab + 2a - 2}{4b+2a-2}} \right)\\
=&O_{\eps}\left(n\right) + \Ohat\left( n^{1+b} \cdot \left( n^{\frac{-3b^2 + ab - b}{3b - a + 1}} \eee^{\frac{(-3b^2 + ab - b)(-2b-a-4)}{(6b-2a+2)(2b+a-1)}} \right) \eee^{\frac{-2b^2 - ab + 2a - 2}{4b+2a-2}} \right)\\
=&O_{\eps}\left(n\right) + \Ohat\left( n^{1+b} \cdot \left( n^{-b} \eee^{\frac{-b(-2b-a-4)}{2(2b+a-1)}} \right) \eee^{\frac{-2b^2 - ab + 2a - 2}{4b+2a-2}} \right)\\
=&O_{\eps}\left(n\right) + \Ohat\left( n \cdot \eee^{\frac{2b^2 + ab + 4b}{4b+2a-2}} \cdot \eee^{\frac{-2b^2 - ab + 2a - 2}{4b+2a-2}} \right)\\
=&O_{\eps}\left(n\right) + \Ohat\left( n \eee \right),
\end{align*}
completing the edge bound.
\end{enumerate}

Plugging in $a=b=2/3$ gives error of
$$\beta = \Ohat\left( n^{3/7} \eee^{-9/7} \right),$$
exactly as in the previous section.
Plugging in $a=1/2, b=1$ for the other distance preserver bound in \cite{CE06} gives
$$\beta = \Ohat\left( n^{3/7} \eee^{-13/14} \right),$$
thus establishing that Theorem \ref{thm:adddspan} holds unconditionally when $\eee = O(1)$.
We can also apply our bounds in Theorem \ref{thm:goodddps}, although we need to be a little more careful here since the respective settings of $a=1, b=1/3$ and 
$a=b=2/3$ only hold on some range of parameters.
The easiest way forward is to proceed through the previous proof, setting all parameters using \emph{both} possible settings of $a,b$.
For example, a bottlenecked cluster would be one that satisfies
$$\left| B_{=}(v, r^*)\right| \le \left| B(v, r^*) \right|^{\frac{1-a}{2b}} \eee^{\frac{1}{2b}}$$
for \emph{both} choices $a=1, b=1/3$ and $a=b=2/3$.
The associated large cluster size bound is then
$$x_r = \Omega\left(\min_{(a=1, b=1/3), (a=b=2/3)} \left\{ r^{\frac{2b}{2b+a-1}} \eee^{\frac{1}{2b+a-1}} \right\}\right),$$
and so on.
In the end, the spanner error is
$$\beta = \Ohat\left(n^{3/7} \eee^{-9/7} + n^{2/3} \eee^{-17/6} \right).$$
The former term dominates in the parameter regime
$$\eee = \Omegahat\left( n^{2/13} \right),$$
while the latter term dominates when
$$\eee = \Ohat\left( n^{2/13} \right).$$
Thus, in particular, Theorem \ref{thm:adddspan} holds unconditionally in the parameter regime $\eee = \Omegahat\left( n^{2/13} \right)$.